\documentclass[fleqn,10pt]{wlscirep}
\usepackage[utf8]{inputenc}
\usepackage[T1]{fontenc}
\usepackage{siunitx}
\usepackage[version=4]{mhchem}
\usepackage[textsize=small]{todonotes}
\setuptodonotes{inline}
\title{Photoacoustic microscopy with meta-optics}

\author[1,2]{Dorian S. H. Brandmüller}
\author[3]{David Grafinger}
\author[1,2]{Robert Nuster}
\author[1]{Andreas Hohenau}
\author[3,4]{Marcus Ossiander}
\author[1,2,*]{Peter Banzer}
\affil[1]{Institute of Physics, University of Graz, Universitätsplatz 5, 8010 Graz, Austria}
\affil[2]{Christian Doppler Laboratory for Structured Matter Based Sensing, Institute of Physics, University of Graz, Universitätsplatz 5, 8010 Graz, Austria}
\affil[3]{Institute for Experimental Physics, Graz University of Technology, Petersgasse 16, 8010 Graz, Austria}
\affil[4]{Harvard John A. Paulson School of Engineering and Applied Sciences, 9 Oxford Street, Cambridge, MA 02138, USA}
\affil[*]{peter.banzer@uni-graz.at}

\begin{abstract}
Recent advances in the miniaturization of optical elements have led to the emergence of novel imaging systems, used for industrial and consumer-based applications. The underlying methods are particularly prevalent in the realms of medical imaging and optical microscopy. Avoiding bulky optical elements can be extremely beneficial to many microscopy modalities, one of which is photoacoustic microscopy. Relying on short, highly focused light pulses that need to be precisely controlled, large and heavy optical elements can often hinder the overall performance of such systems. We propose the utilization of increasingly popular optical elements, so-called meta-optics, in the excitation path of a photoacoustic microscope. The metalenses, which were designed and used for this work, consist of sub-wavelength elements that enable elaborate phase control of incident light and multifunctionality within a single optical element. This allowed us to not only replace common optical elements in the excitation path of the photoacoustic microscope, completely omitting any conventional glass elements, but also to design an adapted lens, increasing the depth of field. With our work, we prove the benefit of meta-optics for photoacoustic microscopy by comparing two different metalenses to a conventional glass lens in simulations as well as experiments. We expect this to be a step into the direction of more advanced meta-optics being utilized in photoacoustic imaging setups.
\end{abstract}
\begin{document}

\flushbottom
\maketitle
\thispagestyle{empty}

\section*{Introduction}
Lately, the miniaturization of optical elements has become more and more prevalent in many areas of research, including medical applications such as optical microscopy and others. This trend demands smaller and smaller lenses, which get increasingly difficult to fabricate using conventional materials. This limitation can, however, be overcome by employing novel nanofabrication techniques to build planar lenses, so-called metalenses\cite{Khorasaninejad2016,Chen2020,Li2019}.
These lenses consist of sub-wavelength building-blocks, so-called meta-atoms, that can be used to manipulate the phase of light element-by-element and thereby shape the wavefront of laser beams. This allows for the design of flat lenses and other optical elements that can be used in various applications such as polarization, phase or light field imaging\cite{Li2019,Kim2021,Lee2020,Chen2021,Zou2020,Lee2020a,Chen2021a,Mousavi2024}.
While most of the previously mentioned works focus on the use of metalenses in imaging and beam shaping applications, there are only a few reports on using those novel optical elements in the field of photoacoustic imaging\cite{Barulin2023,Song2023,Zhao2023}. 

Photoacoustic imaging is a powerful technique that involves a combination of sample excitation with a pulsed light-source and the detection of acoustic waves that are generated inside the sample. This energy conversion effect, which is generally known as the photoacoustic effect is based on the local heating effect eventually resulting in the generation of ultrasound waves, can be exploited for the reconstruction of high-resolution images of the sample and even gain a functional insight into the sample's composition, by exploiting wavelength-dependent absorption characteristics\cite{Beard2011,Park2023,Dean-Ben2017,Xia2014}. In photoacoustic microscopy (PAM) this is most often achieved by scanning a tightly focused laser beam across the sample while detecting the generated ultrasonic waves at each excitation point. Depending on the type of photoacoustic microscope, the resolution is therefore either determined by an acoustic lens that is part of the ultrasound receiver (acoustic resolution PAM, AR-PAM) or by the optical element that is used to focus the excitation laser-beam (optical resolution PAM, OR-PAM)\cite{Jeon2019,Wang2016,Zhu2024}.
One particular field of application for photoacoustic instruments that has become more prominent over time is medical and biological imaging of living organisms and the field of histology. The possibility of achieving a high lateral resolution, while retaining a large depth of field, combined with the ability to gain functional information, such as the oxygenation of blood, makes PAM an ideal tool\cite{Shrestha2020,Paltauf2020,Dean-Ben2017,Dean‐Ben2021}.

PAMs can be implemented in many ways, but one of the difficulties that all of them face is the need to bring comparatively bulky optical and acoustic elements into close proximity to the sample, properly aligned with respect to each other. Reducing the footprint of the focusing elements and reducing their number can help to overcome these limitations. Replacing conventional lenses with metalenses is one way to target this issue, as they can be made microscopically small and flat. This even allows for the direct integration of metalenses into other optical elements, such as optical fibers, giving them an even wider range of flexibility\cite{Hadibrata2021}.

Further, metalenses cannot solely be used as replacements for existing optical elements; they enable far-reaching manipulation of light fields in a matter that is not easily achievable with conventional lenses. Exploiting the arbitrary phase profile that can be imprinted on a beam of light using a metalens allows for structuring the beam in an almost arbitrary fashion, even allowing for reconfigurability, relying only on one element\cite{Yu2014,Khonina2024}. This can be used to create beams or foci with a high depth of field, or beams with tailored intensity, phase and polarization properties, which can be beneficial for certain applications of PAM\cite{Ali2022,Jiang2016}.

In this work, we present the design, simulation, fabrication and, most importantly, experimental characterization of two metalenses that were used in an OR-PAM setup. The first metalens was designed using a basic approach by only utilizing a phase function to focus normally incident light, while the second lens included the phase function of a blazed grating, to separate the focus from the zeroth order and thereby reduces the influence of parasitic light, which is contained in the zeroth order, on the PA signal generation. This platform has the potential to be used in various applications in PAM and is easily adaptable and extendable to more intricate designs, such as multi-focal metalenses or even metalenses with a tunable focal length.

\section*{Results}
\subsection*{Working principle and setup}
We built an OR-PAM with an interchangeable resolution defining optical lens. This yields the opportunity to test various lenses for their resolving power, including metalenses, which were the main target of this work. The excitation light-source for our PAM is a pulsed laser with a central wavelength of \SI{532}{\nano\meter} and a pulse duration of \SI{2}{\nano\second}, operating at a repetition rate of \SI{500}{\hertz}.

\begin{figure}
    \centering
    \includegraphics[width=\textwidth]{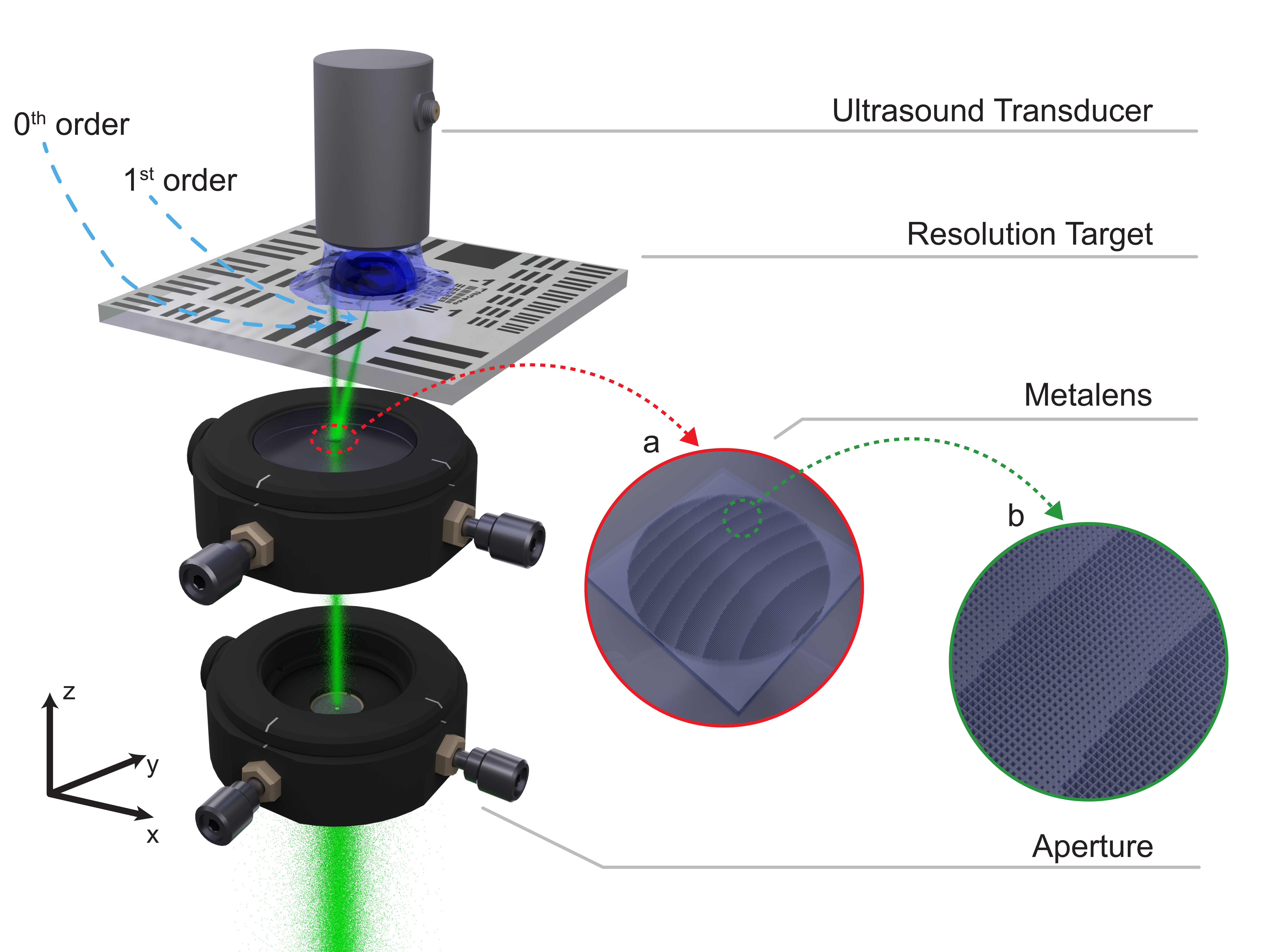}
    \caption{\textbf{Schematic illustration of the experimental setup.}\\After passing through an aperture, the excitation laser beam propagating along the positive z-axis is focused onto the USAF-1951 resolution target by the lens under investigation. The photoacoustic signal is detected by the ultrasound transducer (UT). The insets show detailed views of a metalens schematic including a diffraction grating phase function \textbf{a} and the nanostructures \textbf{b}.}
    \label{fig:setup}
\end{figure}

A schematic illustration of the essential parts of the setup is shown in Fig.~\ref{fig:setup}. The optical components include an aperture with a diameter of \SI{600}{\micro\meter}, to ensure an equal illumination area of both metalenses and the conventional lens as a reference. The lens under investigation is placed close to this aperture, focusing the beam directly onto a USAF-1951 resolution target that is mounted on a magnetically driven xy-stage for easy scanning and a manual high-precision z-stage to aid the overall alignment with respect to the lenses focal plane. A piezoelectric ultrasound transducer (UT), used solely as a receiver, is brought into contact with the sample, using a thin layer of water, ensuring optimal acoustic coupling. The transducer is equipped with an acoustic lens, with a focal spot size of approximately \SI{80}{\micro\meter}. This ensures that only acoustic signals from certain areas are recorded, while the used optical lens stays the resolution defining element, since its focal spot is one order of magnitude smaller. The acoustic lens also ensures that signals from only one of the diffraction orders are recorded when examining metalenses that include the phase of a diffraction grating.

Both the aperture and the lens under investigation are mounted in xy-translation mounts, allowing for precise alignment with respect to the optical axis of the setup. The UT is mounted on a xyz-stage, aiding the alignment of the acoustic to the optical focus and therefore to the origin of the acoustic wave generation.\\
The signals collected by the UT are amplified and recorded using a digital storage oscilloscope. Triggering accuracy was ensured by a photodiode that was placed close to the beam path, recording occurring stray light of the excitation laser beam.

\subsection*{Metalens design, fabrication and optical characterization}
Many state-of-the-art metalenses consist of materials that need highly specialized fabrication techniques to reach their full potential, such as \ce{TiO2}\cite{Khorasaninejad2016, Chen2017}, \ce{Si3N4}\cite{Zhan2016} or \ce{Si}\cite{Liang2018} (used in the NIR and visible range). The use of these specific transparent dielectrics can be circumvented by using resists that are already part of the fabrication process in electron-beam lithography as dielectric materials. This allows for a substantial reduction in the number of processing steps, while retaining a good optical performance and flexibility in the design \cite{Andren2020}.

Most basic metalenses are designed to mimic the focusing properties of conventional lenses as close as possible. This is achieved by creating a focusing phase profile across the metalens' surface by strategically placing nanostructures of different shapes and sizes on a substrate\cite{Aieta2012,Pan2022}. Given a focal length $f$ and a wavelength $\lambda$, the lenses' phase profile $\varphi_L(x, y)$ can be described by the following equation\cite{Hecht2016,Pan2022}
\begin{equation}\label{eq:phase_profile}
\varphi_L(x, y) = \dfrac{2\pi}{\lambda}\left(f - \sqrt{(x^2+y^2)+f^2}\right).
\end{equation}

The original metalens (OM) let a non-negligible amount of unaltered (unfocused) light pass. To avoid any unwanted signals from this background, we additionally designed a metalens with an added blazed grating phase profile (GM). This grating phase profile deflected focused light to its first diffraction order, i.e. away from the zeroth order (see also Fig.~\ref{fig:setup}), and thus separated altered from unaltered light. As we will show later, this leads to an increase in the contrast of the resulting signals and therefore to an overall improvement in the PAM image.

The required grating, for normal incidence, was calculated using the well-known equation\cite{Hecht2016}
\begin{equation}\label{eq:grating}
m\lambda = d \sin(\theta),
\end{equation}
where $m$ is the diffraction order, $\lambda$ the wavelength of the light, $d$ the grating period, and $\theta$ the angle of diffraction.\\
The spatial separation $D$ of the focus from the zeroth order in the focal plane can be calculated as
\begin{equation}\label{eq:separation}
    D = f\tan(\theta).
\end{equation}
The full phase profile $\varphi_D$ to be encoded on the metalens is then given by the sum of the focusing phase profile and the grating phase profile
\begin{equation}\label{eq:full_profile}
    \varphi_D(x, y) = \varphi_L(x, y) + \varphi_G(x, y).
\end{equation}
Here, $\varphi_G(x, y) = 2\pi\left(\tfrac{x}{d_x}+\tfrac{y}{d_y}\right)$, where $d_x$ and $d_y$ are the blazed grating periods in the x and y direction, respectively. 

\begin{figure}
    \includegraphics[width=.5\textwidth]{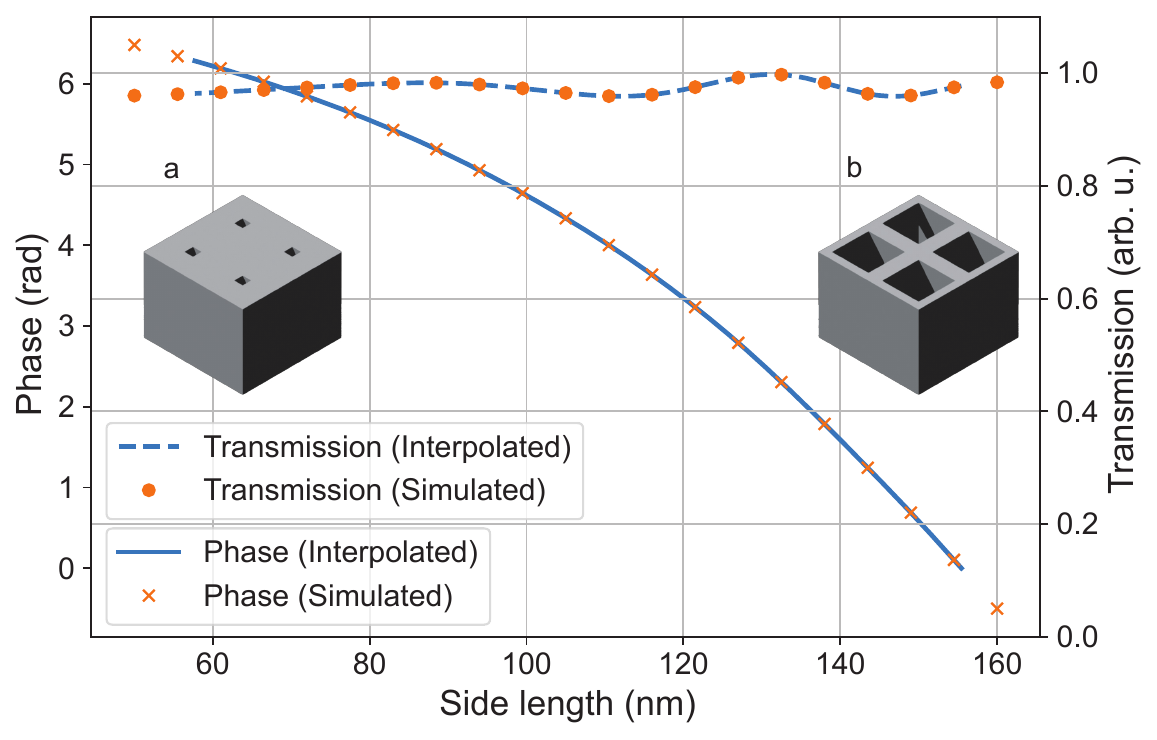}
    \caption{\textbf{Result of FDTD simulations for individual unit cell geometries.}\\Showing the relation of phase shift and transmission of square nanohole arrays as a function of the side lengths of individual holes. The insets show four unit cells each, one group having holes with a short side length \textbf{a} and one holes with a long side length \textbf{b}. The full arrays were realized by employing periodic boundary conditions.}
    \label{fig:phase_vs_radius}
\end{figure}

To identify the relation between size of our individual nanostructures and the respective phase shift, finite-difference time-domain (FDTD) simulations were carried out. These simulations model an infinite array of cuboid shaped holes, set in a PMMA layer with a thickness of \SI{1.6}{\micro\meter}. They were repeated for various hole sizes. As can be seen in Fig.~\ref{fig:phase_vs_radius}, the phase shift induced by the modeled structures is highly dependent on their lateral size. Another important result of these simulations is the dependence of the transmission on the size of the nanostructures, since the overall transmitivity should stay constant and be as high as possible. The simulated structures feature a transmitivity close to 1 (see Fig.\ref{fig:phase_vs_radius}). The simulated results were interpolated using a cubic spline to allow for more precise control over the phase profile of the metalens. A more detailed description of the simulation setup can be found in the methods section.

The results of these simulations showed that side lengths of our nanoholes from \SI{60}{\nano\meter} to \SI{155}{\nano\meter}, were sufficient to fully cover the whole phase range from $0 \text{ to } 2\pi$. These results, together with Eq.~\ref{eq:phase_profile}, allowed us to design a metalens with a focal length of $f = \SI{3}{\milli\meter}$ and a diameter of \SI{500}{\micro\meter}, for a wavelength of $\lambda = \SI{532}{\nano\meter}$.\\
The second metalens we designed included the phase of a diffraction grating, leading to an overall phase profile according to Eq.~\ref{eq:full_profile}. The focal length was set to \SI{5}{\milli\meter} and the grating period was chosen as $d_x = \SI{6.1}{\micro\meter}$, the grating term in the y-direction was ommited. This lead to diffraction angle of $\theta = \SI{5}{\degree}$. Which in turn, according to Eq.~\ref{eq:separation}, resulted in a spatial separation of $D = \SI{437}{\micro\meter}$ between the focus in the first and the zeroth order, a sufficient distance to avoid any interference. A schematic depiction of a smaller version of this metalens, and its placement in the PAM, can be seen in the insets of Fig.~\ref{fig:setup}.

\begin{figure}
    \centering
    \includegraphics[width=\textwidth]{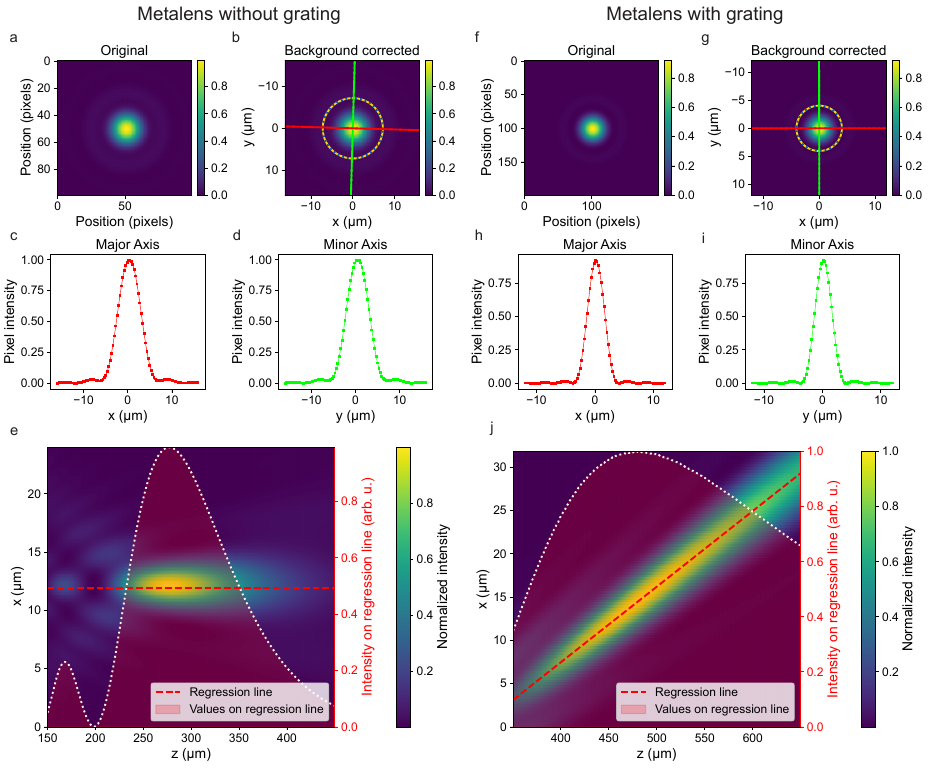}
    \caption{\textbf{Simulated focal fields of the metalenses.}\\Results of the FDTD simulations for the focal fields of the down-scaled metalenses. Subfigures \textbf{a-e} show the results for the focal field of the metalens without a grating, while subfigures \textbf{f-j} show the results for the focal field of the metalens with a grating. Subfigures \textbf{a} and \textbf{f} show the images of the focal plane as they were simulated, subfigures \textbf{b} and \textbf{g} depict background corrected versions of these. They were scaled to the actual dimensions of the focal field and indications of the profiles that were taken, were added, which are shown in subfigures \textbf{c, d, h} and \textbf{i}. Finally, subfigures \textbf{e} and \textbf{j} show the profiles along the optical axis, including a regression line to indicate the depth of field.}
    \label{fig:focal_fields_simulation}
\end{figure}

Further simulations were carried out to ensure that the phase profile of the metalenses was as close to the desired one as possible and that the resulting focus was within the design specification. Since full simulations of metalenses are computationally expensive, we decided to perform the calculations for down-scaled versions of the metalenses, with one-tenth of the size of the manufactured ones, while keeping the same numerical aperture. The resulting simulated focal fields can be seen in Fig.~\ref{fig:focal_fields_simulation}.\\
These results show several profiles taken through the focal fields of the metalenses. One depicts the intensity of the light field in the laterally oriented focal plane, while the other displays the xz-section at position $y = 0$, indicating the light intensity distribution along the propagation direction. Additionally, one-dimensional profiles through the aforementioned sections are shown. To get an idea of the behavior of the depth of field for these lenses, the intensity along a regression line through the focal spot is overlaid with the profiles \textbf{e} and \textbf{j}. This clearly shows a strong elongation of the focal spot in the axial direction for the GM.

\begin{figure}
    \includegraphics[width=.5\textwidth]{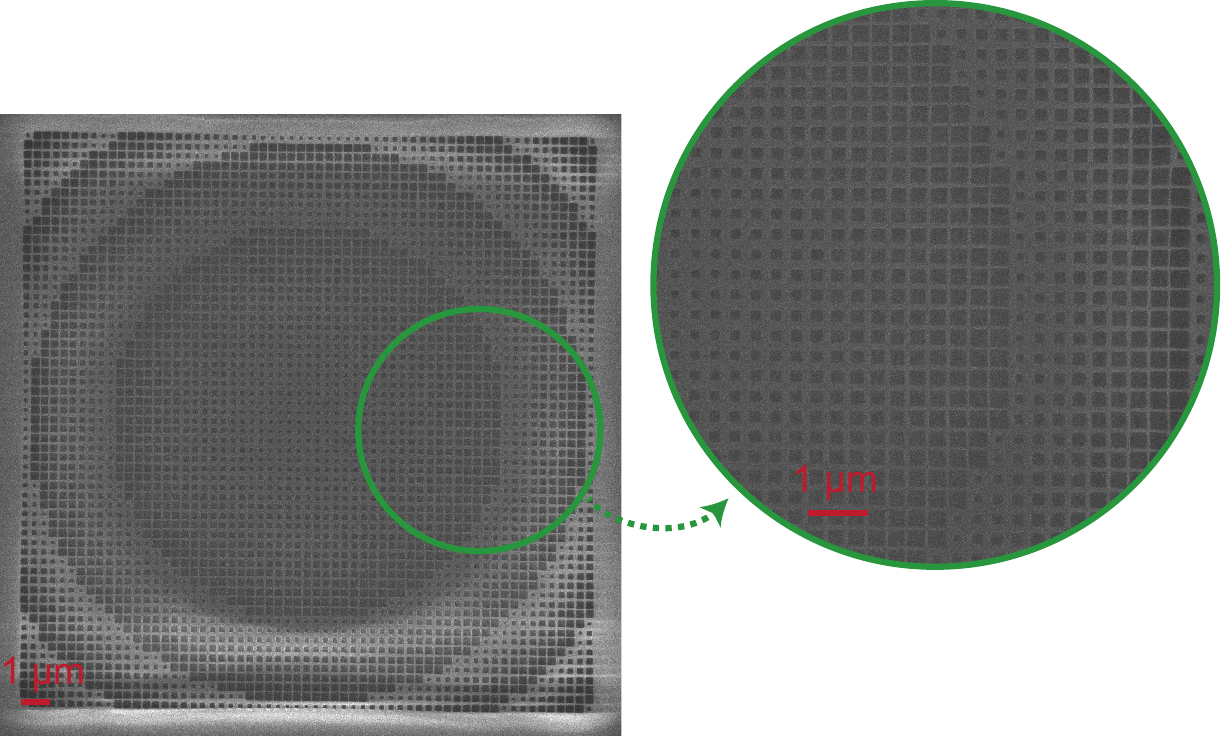}
    \caption{\textbf{Scanning electron microscope images of a fabricated metalens.}\\
    SEM image of the metalens with a diameter of \SI{25}{\micro\meter}, designed without a grating. The inset shows a clear view of the individual unit cells, each sized at \SI{350}{\nano\meter}, containing cuboid-shaped holes with different side-lengths.}
    \label{fig:metalens_sem}
\end{figure}

By using a single-step electron-beam lithography process, we were able to fabricate the designed metalenses with a high accuracy. To evaluate the quality of the produced structures, selected samples were investigated using a scanning electron microscope (SEM). The exemplary SEM images of a small metalens, fabricated with the same nanostructure sizes that were used for the final metalenses mentioned above, are shown in Fig.~\ref{fig:metalens_sem}.

\begin{figure}
    \centering
    \includegraphics[width=1\textwidth]{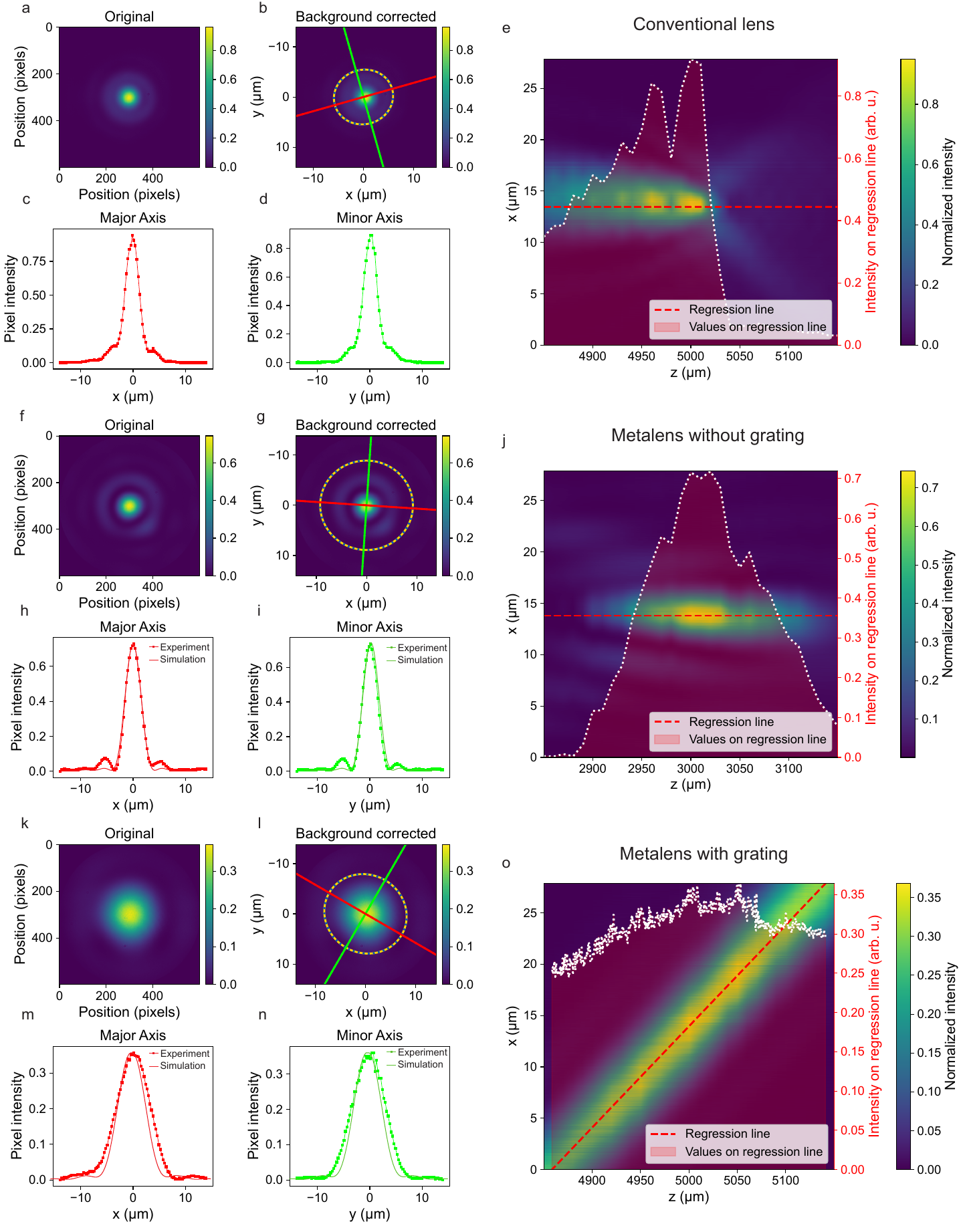}
    \caption{\textbf{Experimentally measured focal fields of the investigated lenses.}\\Experimental results for a conventional glass lens (a-e), the metalens without additional grating phase \textbf{(f-j)}, and for the metalens with a superposed grating phase \textbf{(k-o)}.
    }
    \label{fig:focal_fields_experiment}
\end{figure}

Before these lenses were used in the PAM, their optical performance was tested and compared to a conventional glass lens with a focal length of \SI{5}{\milli\meter}. Regarding the details about the setup used to characterize the focal field of the lenses, we refer to the methods and materials section. In essence, it consists of a collimated laser beam generated by a supercontinuum source, an acousto-optical tunable filter (AOTF) to select the design wavelength of the lenses to be characterized, the same aperture as in the PAM setup and a microscope objective to get 100$\times$ magnified image of the focal spot onto a camera. The latter ensures a precise analysis of the lateral field distribution. Several images were recorded at different distances from the lens under investigation to get a full three-dimensional intensity map of the focal field. The results are shown in Fig.~\ref{fig:focal_fields_experiment}.

The simulation results that are shown in Fig.~\ref{fig:focal_fields_simulation} and the experimental results in Fig.~\ref{fig:focal_fields_experiment} are in good agreement. In addition to the SEM images of the fabricated metalens (see Fig.~\ref{fig:metalens_sem}), this is another strong indicator of the high quality of the fabricated metalenses. For a graphical comparison, the simulated profiles for the minor and major axes of the focal spots were overlaid with the experimental results. The full width at half maxima (FWHMa) of the focal spots were calculated by fitting Bessel functions to the intensity profiles and are summarized in Table~\ref{tab:lenses}.
\begin{table}
    \centering
    \begin{tabular}{| l | S[table-format=1.3] | S[table-format=1.1] | S[table-format=1.1] | S[table-format=1.1] | S[table-format=2.2] |}
        \hline
        {Lens}                    & {$\mathit{NA}$} & {$f$ (\si{\milli\meter})} & {FWHM\textsubscript{s} (\si{\micro\meter})} & {FWHM\textsubscript{e} (\si{\micro\meter})} & {PAM\textsubscript{res} (\si{\micro\meter})} \\
        \hline
        Conventional glass lens   & 0.060           & 5                   &                                       & 3.1                                      & 5.52 \\
        Metalens without grating (OM) & 0.083           & 3                   & 3.3                                   & 3.4                                      & 11.05 \\
        Metalens with grating (GM)    & 0.050           & 5                   & 5.7                                   & 7.3                                      & 6.22 \\
        \hline
    \end{tabular}
    \caption{\textbf{Results of the  lens characterization.}\\ The numerical aperture ($\mathit{NA}$), the focal length ($f$), the FWHM of the simulated (FWHM\textsubscript{s}) and the experimental (FWHM\textsubscript{e}) focal spot, as well as the lateral resolution of the PAM in the focal plane (PAM\textsubscript{res}) are shown.}
    \label{tab:lenses}
\end{table}

The results unambiguously prove that our fabricated metalenses can focus light to a spot size that is comparable to that of a conventional glass lens, with a slightly worse lateral resolution.\\
The GM showed a larger focal spot size, which can be attributed to the additional non-centrosymmetric phase profile that was introduced. However, a strong elongation of the focal spot in the axial direction was observed, which is beneficial for some applications of PAMs\cite{Ali2022, Jiang2016}. One area of research where these longitudinally elongated foci show their merits is histological imaging, where a large depth of focus allows for high-resolution imaging, despite some irregularities in the sample surface\cite{Zhao2023, Song2023}.
The separation of the focus through the splitting into different diffraction orders also allowed for an easy measurement of the metalenses efficiency, as the individual parts were measured using a knife edge for blocking certain parts of the beam. The calculation of the efficiency was then carried out by a simple comparison of the intensities of the blocked and unblocked beams, following equation
\begin{equation}\label{eq:efficiency}
    I_{1} = I_{total} - \sum_{i \neq 1}^{n}I_i,
\end{equation}
where $I_{total}$ is the total intensity of the beam and $I_i$ are the intensities of the individual diffraction orders, while $i = 1$ represents the focused beam used for PAM measurements. The efficiency of the metalens was then calculated as the ratio of the intensity of the focused beam to the total intensity of the beam $\tfrac{I_1}{I_{total}}$. This results in a measured efficiency of \SI{20}{\percent}, not quite matching the percentages of around \SI{50}{\percent} reported for similar resist-only metalenses\cite{Andren2020}. Considering absorption and scattering losses within the metalens, the efficiency drops to \SI{17}{\percent}. However, considering the high aspect ratio of our nanostructures (up to 1:26) and the low refractive index of our PMMA resist ($n = 1.49$), this result is still within the expected range.\\
The efficiency of the conventional glass lens was measured to be \SI{90}{\percent}, comparing the incoming and transmitted power.

\subsection*{Characterization of the photoacoustic microscope}
\begin{figure}
    \includegraphics[width=0.5\textwidth]{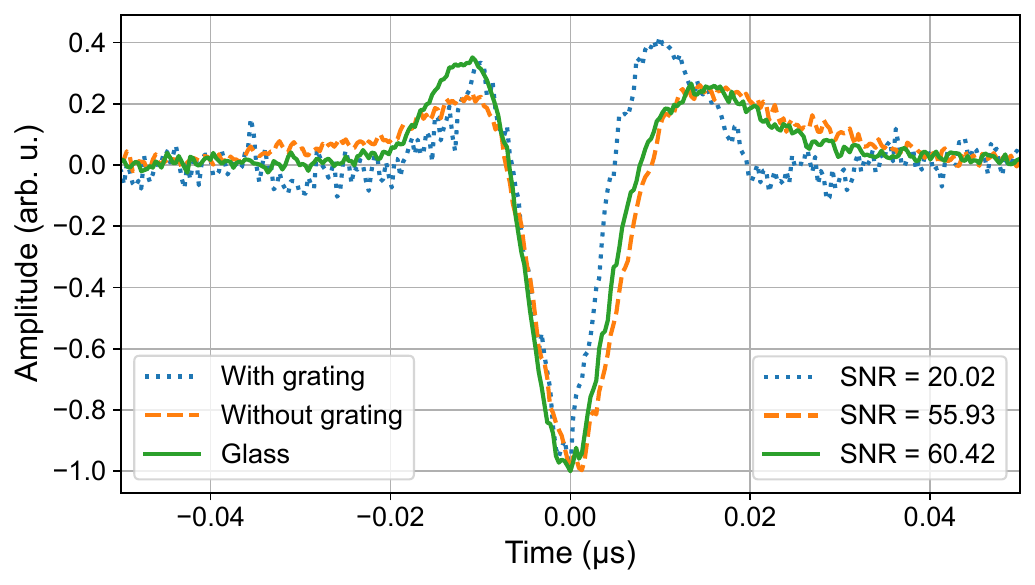}
    \caption{\textbf{Comparison of photoacoustic signals.}\\
    Individual PAM time-signals for all three lenses. The signals were normalized to their negative peak value, the average was subtracted, and the signals were shifted in time to have the minimum at time zero. The signal-to-noise ratio (SNR) is given by the ratio of the minimum to the standard deviation of the signal's noise.}
    \label{fig:signal_comparison}
\end{figure}

To fully characterize the performance of our PAM system with these lenses, they were inserted into the setup (see Fig.~\ref{fig:setup}) and a series of time traces and  2D scans were recorded. These measurements were performed with varying positions of the sample along the optical axis. The individual signals were generated with a highly absorbing material to obtain a sufficiently good signal quality for all lenses, to get an idea of their overall performance and to see how they compare to each other in the time domain. The signals were normalized to their minimum value, the average of the signal was subtracted, and it was shifted in time to have the minimum of the signal at time zero.\\
The average power of the laser was set to \SI{138}{\micro\watt} for the metalenses and to \SI{22.8}{\micro\watt} for the conventional glass lens. This power reduction was necessary to avoid any damage to the sample, since the glass lens had a much higher focusing efficiency than the metalenses. The power was measured directly after the aperture for each measurement to ensure that absorption losses did not influence the results and to determine the focussing efficiency by relating the measured values to the generated PAM signal amplitudes. The individual signals that were captured can be seen in Fig.~\ref{fig:signal_comparison}.

\begin{figure}
    \centering
    \includegraphics[width=\textwidth]{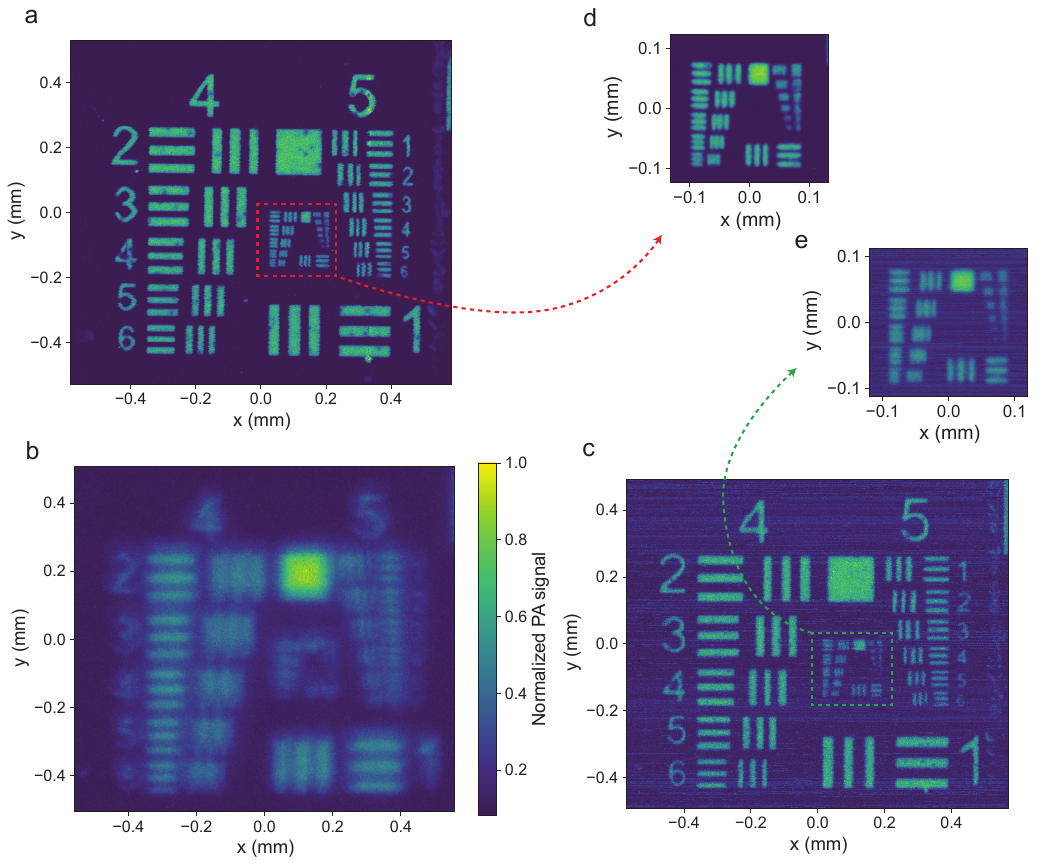}
    \caption{\textbf{Resolution target scans.}\\MAPs of the scans of the USAF-1951 resolution target, taken with \textbf{a} the conventional glass lens, \textbf{b} the metalens without a grating phase and \textbf{c} the metalens with a grating phase. The subfigures \textbf{d} and \textbf{e} show the zoomed-in views of the groups 6 and 7, for the glass lens and the metalens with a grating phase, respectively.}
    \label{fig:resolution_target}
\end{figure}

Comparing the signals reveals that all the temporal PAM signals recorded with the three different lenses feature a similar shape. The form being characteristic for acoustic waves generated by pulsed laser radiation in the acoustic focal plane and measured with a piezoelectric UT with a limited bandwidth.  Moreover, the signals are almost of the same width after being normalized. This already indicates the suitability of metalenses for PAM. In addition, the signal-to-noise ratio (SNR) of the metalenses is noticeably lower than that of the conventional glass lens. This can be attributed to the lower focusing efficiency of the metalenses, which is five-times less. The difference in SNR between the OM and the GM stems from the unperturbed light intensity that is filtered by the added grating phase function of the GM. Although this intensity leads to a higher SNR for the OM, it is not beneficial to the overall imaging results, as can be seen in Fig.~\ref{fig:resolution_target}.
The investigation of the lateral resolution was carried out using a USAF-1951 resolution target. The target was placed in the focal plane of the lens under investigation and scanned in the xy-plane. The maximum amplitude projections (MAPs) of these scans are shown in Fig.~\ref{fig:resolution_target}.

With our reference lens, we were able to resolve element~4 of group~6, which is equivalent to a resolution of \SI{5.52}{\micro\meter}. This served as a baseline for the other measurements.
The metalens without a grating phase was able to resolve up to element~4 of group~5, hence featuring a lateral resolution of \SI{11.05}{\micro\meter}. The metalens with a grating phase was able to resolve up to element~3 of group~6, i.e. a lateral resolution of \SI{6.2}{\micro\meter}.
While having slightly different numerical apertures, due to the different focal lengths of the lenses and the aperture size (see Tab.~\ref{tab:lenses}), the resolutions are comparable. The resolution of the metalenses is slightly worse than that of the reference lens, which was expected, when considering the FWHM of the focal spots of the lenses. Comparing the two metalenses shows that even though the OM has a smaller focal spot size, its lateral resolution is measurably lower than for the GM. This stems from the signal that is generated by the light field that is not being focused, which is still sufficiently strong to generate sound waves that are detected by the US. Consequently, the OM exhibits reduced contrast, as can be seen in Fig.~\ref{fig:resolution_target}~b. Hence, the added grating phase function clearly shows its benefits, when it comes to the lateral resolution of the PAM in the plane of focus.

\begin{figure}
    \centering
    \includegraphics[width=\textwidth]{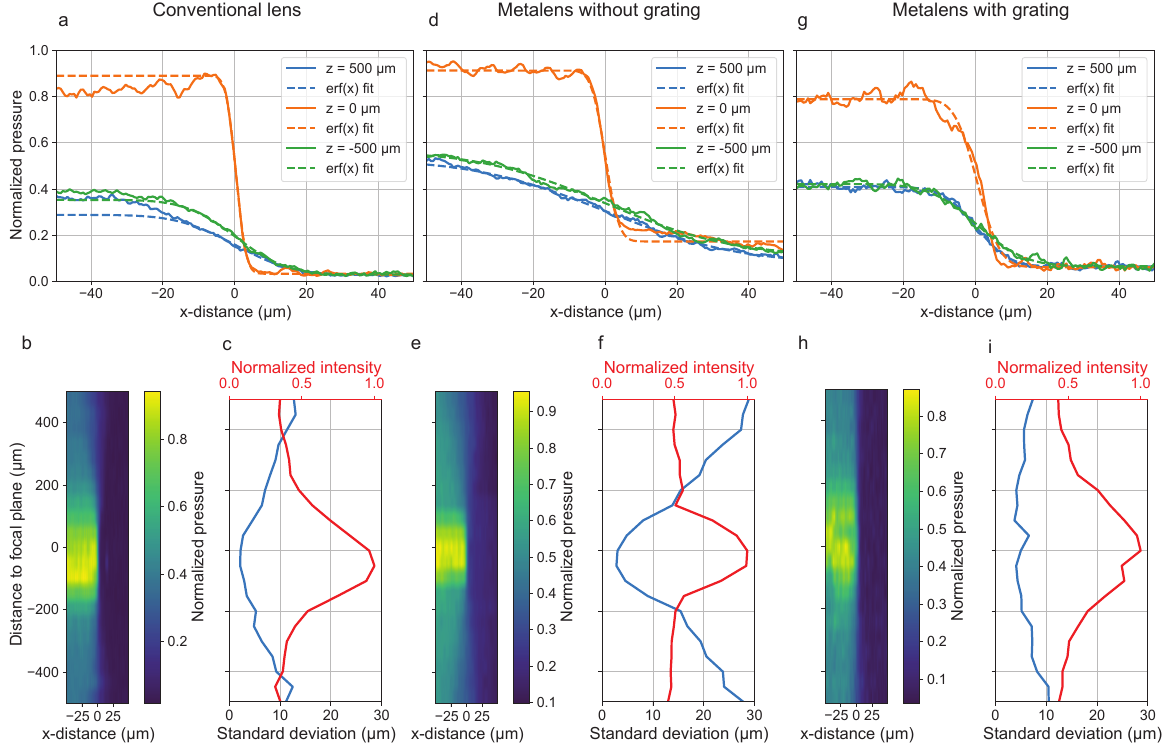}
    \caption{\textbf{Depth of field measurements.}\\MAPs of line scans over a sharp edge, taken with \textbf{a-c} the conventional glass lens, \textbf{d-f} the metalens without a grating phase and \textbf{g-i} the metalens with a grating phase. The z-distance is measured from the focal plane, where a positive distance indicates a distance further away from the lens and vice versa. Subplots \textbf{a, d} and \textbf{g} show line scans at three distances (\SI{-500}{\micro\meter}, \SI{0}{\micro\meter}, \SI{500}{\micro\meter}) from the focal plane, including fitted Gaussian error functions (see Eq.~\ref{eq:gef}). Subplots \textbf{b, e} and \textbf{h} depict the line scans at all distances, while subplots \textbf{c, f} and \textbf{i} show the standard deviation of the fitted Gaussian error functions and the normalized intensity of the line scans.}
    \label{fig:depth_of_field}
\end{figure}

To characterize the different lenses with respect to their depth of field, we performed one-dimensional scans over a region with several sharp edges on the resolution target. This was done for increasing distances from the focal plane for each lens. Additionally, these measured lines were fitted with Gaussian error functions
\begin{equation}
    \label{eq:gef}
    f(x, \sigma) = \text{erf}\left(\dfrac{x}{\sqrt{2}\sigma}\right).
\end{equation}
Here, $x$ is the distance to the edge and $\sigma$ is the standard deviation of the error function. The underlying assumption of this method is a radially symmetric Gaussian point spread function of the PAM system, with a standard deviation $\sigma$. The results of these measurements are shown in Fig.~\ref{fig:depth_of_field}. $\sigma$ can be used as a measure of the lateral PAM resolution.

The high lateral resolution of the conventional glass lens was once more confirmed by these measurements, indicated by the sharp edge, and therefore low $\sigma$, of all measured lines. But after a distance of \SI{500}{\micro\meter} from the focal plane, the signal strength starts to significantly weaken, which is unfavorable for imaging samples requiring a large depth of field\cite{Zhao2023}.\\
A similar trend can be seen for the OM (see Fig.~\ref{fig:depth_of_field}~\textbf{e} and \textbf{f}). Even though, the lateral resolution is quite good in the focal plane, it worsens quickly for distances further away from the focal plane.\\
The GM, on the other hand, shows a much better performance, in both the lateral resolution and the signal strength over a large region of distances from the focal plane. Even for distances of up to \SI{500}{\micro\meter} from the focal plane, the resolution and signal strength stays high, as expected from the focal field (see Fig.~\ref{fig:focal_fields_experiment}). Overall, the resist-only metalens with the added diffraction grating phase behaved as designed and opens up the possibility for high-resolution imaging over a large depth of field, with a single, easy-to-fabricate optical element.

\section*{Discussion}
We designed and fabricated two metalenses for the excitation of acoustic waves in a PAM. While we chose a purely focusing phase profile for the first metalens, we added a blazed grating phase function to the second one. This grating turned out to be beneficial with respect to the recorded images, as it allowed for a higher contrast and an improved depth of field, proven by simulations of the focal field as well as with experimental results.

The metalenses were fabricated using solely PMMA resist, commonly used in electron-beam lithography. This made the fabrication comparatively straightforward by omitting several processing steps that are usually needed when building metalenses\cite{Wang2022}. The structures were of a high quality, as was confirmed by SEM images of the fabricated lenses. A comparison to a conventional glass lens in our setup also showed the strong performance of the metalenses in the focal volume and the superior performance of the metalens with grating at large distances from the focal plane.\\
While metalens-like structures have been proposed for and used in photoacoustics before\cite{Song2023,Zhao2023}, we were able to experimentally show for the first time, that metalenses can be used in PAMs without the need for additional optical elements, such as glass lenses or objectives. Furthermore, we have shown an extension of the depth of field of the PAM, without introducing additional optical elements, by simply adapting the design of the used metalens. This opens up new possibilities for PAMs, especially in the field of histological imaging, where a large depth of field is crucial for high-resolution imaging\cite{Ali2022,Jiang2016}.

The results of this work show that metalenses are a viable alternative to conventional glass lenses in PAMs, particularly for certain applications, such as histological imaging. They can be used to achieve similar resolution, while also offering the possibility of an extended depth of field and other beam shaping possibilities. This makes them a valuable tool for the field of photoacoustic microscopy, especially in applications where flat and miniaturized optics are beneficial.

\section*{Materials and methods}

\subsection*{Numerical simulations}
To calculate the properties of arrays of single size nanostructures that are shown in Fig.~\ref{fig:phase_vs_radius}, we used a finite difference time domain (FDTD) based commercial software, Lumerical by Ansys Ltd. A layer system was implemented, consisting of a BK7-glass layer and a PMMA layer with a thickness of \SI{1.6}{\micro\meter}. The system was placed in air. To mimic an infinitely extended array of the nanostructures under study, periodic boundary conditions were used along the grid and perfectly matched layers (PML) along the optical axis. The structures are illuminated with a linearly polarized monochromatic plane wave source under normal incidence from the glass side.

This process was repeated for 21 different side lengths of the nanoholes, ranging from \SI{50}{\nano\meter} to \SI{160}{\nano\meter}, recording the phase shift and transmission of the array for each simulation.\\
After fitting the results with a cubic spline, a region of nanoholes was selected that covered the full phase range from $0 \text{ to } 2\pi$, while keeping the transmission close to 1. Care was taken to choose aspect ratios that were easy to fabricate.

To simulate the focal fields of the whole metalenses, we used the same software. Placing nanoholes in the PMMA layer according to Eq.~\ref{eq:phase_profile} and Eq.~\ref{eq:full_profile}, and replacing the periodic boundaries with PMLs, we were able to simulate the focal fields of down-scaled metalenses.

\subsection*{Metalens fabrication}
After performing a general preparation process for the electron-beam lithography (EBL) on the glass substrates, the \SI{1.6}{\micro\meter} thick PMMA layer was spin-coated onto it. Afterward, the sample was exposed with an \SI{30}{\kilo\volt} electron-beam using a clearing dose of \SI{240}{\micro\coulomb\per\cm^2}. The designs used for the metalenses were proximity corrected using the software Beamfox by Beamfox Technologies ApS. The exposure was performed using an eLine Plus system from RAITH~GmbH.\\
After the exposure, the samples were developed and rinsed. No further processing steps were required.

\subsection*{Optical characterization}
We optically characterized the lenses by measuring the focal fields using a microscope setup. A super continuum-source (SuperK Extreme NKT Photonics A/S) combined with a tunable spectral filter was used to generate a collimated narrow bandwidth beam of light. This beam was first cut by an aperture and then focused by the lens under investigation. The generated focal field was then imaged onto a camera using a microscope objective with a numerical aperture of 0.8 at 100$\times$ magnification. To record the complete 3D intensity, images were captured while scanning the objective along the optical axis, close to the focal plane, with a micrometer precision linear stage. 

The efficiency of the metalens with the grating was easily measured by blocking parts of the beam and measuring the power with a power meter, as described in the results section (see also Eq.~\ref{eq:efficiency}).

\subsection*{Photoacoustic microscope characterization}
The general setup of the OR-PAM is described in the results section and can be seen in Fig.~\ref{fig:setup}. It is equiped with a frequency doubled pulsed Nd:YAG light source (BrightSolutions) prividing laser pulses with a central wavelength of \SI{532}{\nano\meter} and a pulse duration of \SI{2}{\nano\second}.\\ 
The  generated photoacoustic signals were captured using a piezoelectric UT, with a central frequency of \SI{50}{\mega\hertz} and a bandwidth of \SI{80}{\percent} (V214-BB-RM Olympus NDT Inc.). In addition, in front of the UT an acoustic lens was attached with \SI{9}{\milli\meter} focal length and acoustic detection aperture of $NA_{ac} = \frac{D}{2f_{ac}} = \SI{0.35}{}$. To increase the amplitude, the signals were amplified using two cascaded voltage amplifiers (Mini-Circuits ZFL-500LN+, \SI{28}{\dB}), the recording was done using a digital storage oscilloscope (Tektronix MSO44). The precise movement of the sample was ensured by two magnetically driven stages (V-408 PIMag Physik Instrumente GmbH \& Co. KG). All components were synchronized by a Python script, controlling the oscilloscope and the stages.
\newpage
\section*{Data availability}
The data that support the findings of this study are available from the corresponding author upon reasonable request.

\section*{Acknowledgements}
The financial support by the Austrian Federal Ministry of Labor and Economy, the National Foundation for Research, Technology and Development and the Christian Doppler Research Association is gratefully acknowledged. The authors thank Jörg Eismann for providing elements of the embedded graphics, Scott Prahl for his work on the Python module \texttt{laserbeamsize}\cite{Prahl2023} and Lasse Frølich from Beamfox Technologies for his support with the beam proximity correction.
M.O. acknowledges funding from the European Union (grant agreement 101076933 EUVORAM). The views and opinions expressed are, however, those of the author(s) only and do not necessarily reflect those of the European Union or the European Research Council Executive Agency. Neither the European Union nor the granting authority can be held responsible for them.
\bibliography{bibliography}
\end{document}